\documentstyle[aps]{revtex}

\begin{document}
\title{Deterministic Secure Direct Communication Using Mixed State}
\author{Qing-yu Cai}
\address{Laboratory of Magentic Resonance and Atom and Molecular Physics, Wuhan\\
Institute of Physics and Mathematics, The Chinese Academy of Science, Wuhan,%
\\
430071,People's Republic of China}
\maketitle

\begin{abstract}
We show an $improved$ $ping-pong$ protocol which is based on the protocol
showed by Kim Bostrom and Timo Felbinger [Phys. Rev. Lett. 89, 187902
(2002)]. We show that our protocol is asymptotically secure key distribution
and quasisecure direct communication using a mixed state. This protocol can
be carried out with great efficiency and speed with today's technology, e.g.
single photon source and linear optical elements technology.

\begin{description}
\item  PACS numbers: 03.67.Hk, 03.65.-a
\end{description}
\end{abstract}

Quantum key distribution is a brilliant application of quantum mechanics. We
consider that quantum channel is secure because quantum physics establishes
a set of negative rules stating things that cannot be done: (1) One cannot
take a measurement without perturbing the system. (2) One cannot determine
simultaneously the position and the momentum of a particle with arbitrarily
high accuracy. (3) One cannot simultaneously measure the polarization of a
photon in the vertical-horizontal basis and simultaneously in the diagonal
basis. (4) One cannot draw pictures of individual quantum processes. (5) One
cannot duplicate an unknown quantum state. These characters of quantum
physics let us have ability to exchange information securely. The motive
that we build quantum-mechanical communications channels is not only to
transmit information securely without being eavesdropped on but also to
transmit information more efficiently.

In the Bennett-brassard (BB84) protocol [1], Alice send a key qubit to Bob,
which is prepared in one of two conjugate bases. Bob measures the qubit in
one of the two base. The eavesdropper Eve does not know the basis choose by
Alice, so she cannot obtain information about the key without a detectable
disturbance. But this type of cryptographic schemes are usually non$\det $er$%
\min $istic [2,3]. A secure direct communication has been presented in
reference [4], which allows encoding only after a final transmission of
classical information. Recently, K. Bostrom and T. Felbinger presented a $%
ping-pong$ $protocol$, which allows for deterministic communication using
entanglement [5]. In this paper, we will give an improved $ping-pong$ $%
protocol$ using single photon in the mixed states.

It is well known that we can prepare a photon in states $\{|0>,|1>\}$ or $%
\{|\varphi _{0}>,|\varphi _{1}>\}$in its polarization degree of freedom,
where 
\begin{eqnarray}
|\varphi _{0} &>&=\frac{1}{\sqrt{2}}(|0>+|1>),\text{ } \\
|\varphi _{1} &>&=\frac{1}{\sqrt{2}}(|0>-|1>).
\end{eqnarray}
Denote that $i\sigma _{y}=|0><1|-|1><0|$, it can be obtained: 
\begin{equation}
i\sigma _{y}|0>=-|1>,\text{ }i\sigma _{y}|1>=|0>
\end{equation}
and 
\begin{equation}
i\sigma _{y}|\varphi _{0}>=|\varphi _{1}>,\text{ }i\sigma _{y}|\varphi
_{1}>=-|\varphi _{0}>.
\end{equation}
Suppose Alice want to obtain some classical information. First she sends a
photon which is prepared in a mixed states $|\psi >:$%
\begin{equation}
\rho =|\psi ><\psi |=\frac{1}{2}|0><0|+\frac{1}{2}|\varphi _{0}><\varphi
_{0}|
\end{equation}
to Bob. That is, Alice selects state $|0>$ or $|\varphi ^{+}>$ randomly with
the probability $p=\frac{1}{2}$ every time. Bob decides either to perform
the operation $i\sigma _{y}$ on the travel qubit to encode the information
'1' or do nothing, i.e., to perform the operation $I$ to encode the
information '0'. Then Bob sends the travel qubit back to Alice. Alice
performs a measurement on this back photon to draw the information Bob
encoded. This is a $ping-pong$ protocol [5]. In this protocol, there are two
communication modes, '$message$ $mode$' and '$control$ $mode$' (see Figs.1
and 2.). By default, Bob and Alice are in message mode. The communication is
described above. With probability $c$, Bob switches the message mode to
control mode. After Bob obtained a photon, instead of performing his
operation on the travel qubit, Bob use the public channel to exchanges
information about the basis Alice used. Then Bob performs a measurement in
the basis $B_{0}=\{|0>,|1>\}$ or $B_{1}=\{|\varphi _{0}>,|\varphi _{1}>\}$
decided by the information Alice told. Bob sends the measurement result to
Alice. If Alice find the result is the same as she prepared, she let the
communication continue. Else, there is an Eve in line. The communication
stops! This protocol can be described explicitly like this:

1) Alice prepares a photon in the state $|0>$ or $|\varphi _{0}>$ with a
probability $p=\frac{1}{2}$.

2) Alice sends the photon to Bob.

3) Bob receives the travel photon. He decides to the message mode or the
control mode randomly.

4c) $Control$ $mode$. Bob exchanges two bits information with Alice through
public channel. Suppose the photon was prepared in state $|0>$, but Bob
finds that the measurement result is in the state $|1>$ (Or Alice prepared
the photon in the state $|\varphi _{0}>$. However, Bob finds the photon in
the state $|\varphi _{1}>$). There is an Eve in line. The communication
stops. Else, this communication continues (Goto 1).).

4m) $Message$ $mode$. Bob performs an operation to encode classical
information on the photon. He encodes the information '0' by the operation $%
I $, and the information '1' by the operation $i\sigma _{y}$. Then Bob sends
the photon back to Alice. Alice obtains the classical information encoded by
Bob with a measurement in the basis it has been prepared. This communication
continues (Goto 1). ).

5) When all of Bob's information transmitted, this communication stops.

We will show that this protocol is secure below.

Eve is an evil quantum physicist able to build all devices that are allowed
by the laws of quantum mechanics. Her aim is to find out which operation
Alice performs (See Fig.3.). First, we will show that Eve can not get full
information about the states $|\psi >$ that Alice prepared. Here we will use
a very famous theorem, the Holevo theorem [6,7]. It states that: Suppose
Alice prepares a state $\rho _{i}$ where $i=0,...,n$ with a probabilities $%
p_{0},...,p_{n}$. Eve performs a measurement described by POVM elements $%
\{E_{j}\}=\{E_{0},...,E_{m}\}$ on the state, with measurement outcome E. The
Holevo theorem states that for any such measurement Eve may do: 
\begin{equation}
H(A:E)\leq S(\rho )-\sum_{i}p_{i}S(\rho _{i}),
\end{equation}
where 
\begin{equation}
\rho =\sum_{i}p_{i}\rho _{i}.
\end{equation}
S($\rho $) is the Von Neumann entropy. The mutual information H(A:E) of A
and E measures how much information A and E have in common. It is a good
measure of how much information has been gained about A (Alice) from the
measurement outcome E (Eve). We denote 
\begin{equation}
\chi =S(\rho )-\sum_{i}p_{i}S(\rho _{i}).
\end{equation}
The Holevo chi quantity is the upper bound on the accessible information. We
will prove that 
\begin{equation}
\chi (\rho )<H(A),
\end{equation}
where H(A) is Shannon entropy as a function of a probability distribution.
The binary entropy H(A) is [8]: 
\begin{equation}
H(A)=-p\log p-(1-p)\log (1-p).
\end{equation}
Since either of the mixed states is prepared with equal probability $p=\frac{%
1}{2}$, then we have 
\begin{equation}
H(A)=1.
\end{equation}
The upper bound on the accessible information Eve can gain about the mixed
states is $\chi $. The von Neumann entropy is defined by [8] 
\begin{equation}
S(\rho )=-tr(\rho \log \rho ).
\end{equation}
The matrix of the density operator in the basis $\{|0>,|1>\}$ is 
\begin{equation}
\rho =|\psi ><\psi |=\left[ 
\begin{array}{ll}
\frac{3}{4} & \frac{1}{4} \\ 
\frac{1}{4} & \frac{1}{4}
\end{array}
\right] .
\end{equation}
We can calculate the eigenvalues $\lambda $ of $\rho $: 
\[
\lambda _{1,2}=\frac{1}{2}\pm \frac{\sqrt{2}}{4}. 
\]
Then von Neumann's definition of entropy (10) can be re-expressed 
\begin{equation}
S(\rho )=-\lambda _{1}\log \lambda _{1}-\lambda _{2}\log \lambda _{2}.
\end{equation}
And the von Neumann entropy is zero for a pure state. So we have 
\begin{equation}
\chi =S(\rho )-\sum_{i}p_{i}S(\rho _{i})<H(A),
\end{equation}
which means the law of quantum mechanics forbids Eve to gain full
information about the states $\rho $ which Alice prepared. Therefore, Eve is
not sure which state Alice prepared.

To find out which operation Bob performs, Eve would use all methods that
quantum mechanics laws allowed. The most general quantum operation is a
completely positive map 
\begin{equation}
\varepsilon :S(H_{A})\rightarrow S(H_{A})
\end{equation}
Because of the Stinespring dilation theorem [9], and completely positive map
can be realized by a unitary operation on a larger Hilbert space. If the $%
H_{A}$ has a Hilbert space of $d$ dimensions, then it suffices to model the
ancilla space $H_{E}$ as being in a Hilbert space of no more than $d^{2}$
dimensions. With an ancilla state $|e>\in H_{E}$, and a unitary operation $U$
on $H_{A}\otimes H_{E}$, for all $\rho _{A}\in H_{A}$, we have 
\begin{equation}
\varepsilon (\rho _{A})=Tr_{E}[U(\rho _{A}\otimes |e><e|)U^{\dagger }].
\end{equation}
In order to gain information about Bob's operation, Eve should first perform
the unitary attack operation $U$ on the composed system, then let Bob
perform his coding operation on the travel photon, and finally perform a
measurement. No measurement implies that Eve did not get any information
about the travel photon.

Because the photon is in mixed states, to Eve, the travel photon is prepared
in a state either $|0>$ or $|\varphi _{0}>$ with a probability $p=\frac{1}{2}
$. First, let us suppose the travel photon is in the state $|\varphi _{0}>$.
Because Eve's purpose is to find out which operation Bob performs, she adds
an ancilla in the state $|e>$ and performs the unitary operation $U$ on both
systems, resulting in 
\begin{equation}
|\psi ^{\prime }>=U|\varphi _{0},e>=\alpha |\varphi _{0},e_{0}>+\beta
|\varphi _{1},e_{1}>,
\end{equation}
where $|e_{0}>,|e_{1}>$ are pure ancilla states uniquely determined by $U$,
and $|\alpha |^{2}+|\beta |^{2}=1$. Randomly, Bob selects the control mode
and turns on the public channel. If Eve does not exist, the result of Bob's
measurement will always in the state $|\varphi _{0}>$. With Eve in line, the
detection probability for Eve's attack in a $control$ $mode$ is 
\begin{equation}
d=|\beta |^{2}=1-|\alpha |^{2}.
\end{equation}
Without $control$ $mode$, after Eve's attack operation, the state of the
system becomes 
\begin{equation}
\rho ^{\prime }=|\psi ^{\prime }><\psi ^{\prime }|,
\end{equation}
which can be written in the orthogonal basis $\{|\varphi _{0},e_{0}>,$ $%
|\varphi _{1},e_{1}>\}$ as 
\begin{equation}
\rho ^{\prime }=\left[ 
\begin{array}{ll}
|\alpha |^{2} & \alpha \beta ^{*} \\ 
\alpha ^{*}\beta & |\beta |^{2}
\end{array}
\right] .
\end{equation}
Bob encodes his one bit by applying the operation $I$ or $i\sigma _{y}$ to
the travel photon with the respective probability $p_{0}$ and $p_{1}$. Then,
the state of the qubit photon becomes 
\begin{equation}
\rho ^{\prime \prime }=\left[ 
\begin{array}{ll}
|\alpha |^{2} & \alpha \beta ^{*}(p_{0}-p_{1}) \\ 
\alpha ^{*}\beta (p_{0}-p_{1}) & |\beta |^{2}
\end{array}
\right] .
\end{equation}
The maximal information Eve can gain is decided by the von Neumann entropy, $%
S(\rho ^{\prime \prime })$, described by Eq.(12). We can calculate the
eigenvalues $\lambda $ of $\rho ^{\prime \prime }$: 
\begin{equation}
\lambda _{1,2}=\frac{1}{2}\pm \frac{1}{2}\sqrt{%
1-(4d-4d^{2})[1-(p_{0}-p_{1})^{2}]}.
\end{equation}
When $p_{0}=p_{1}$, the maximal information Eve can gain is surprising equal
to the Shannon entry of a binary channel, 
\begin{equation}
I(d)=H_{bin}(d)=-d\log d-(1-d)\log (1-d),
\end{equation}
which is concavity for any $d$ range 0 to 1 [8]. The function $I(d)$ has a
maximum at $d=1/2$. It is a monotonous function $0\leq d(I)\leq 1/2$, $%
I(d)\in [0,1]$. When the information Eve gains $I(d)>0$, the detection
probability is $d(I)>0$. If Eve want to gain full information about Bob's
operation, the detection probability is $d(I)=1/2$. Assume that Alice sends
state $|0>$ rather than $|\varphi _{0}>$, the same result we can gain [5].

Suppose the probability of $control$ $mode$ in every protocol run is $c$.
The effective transmission rate is $1-c$. The asymptotic probability after $n
$ bits transmitted that Eve is not detected becomes 
\begin{equation}
p_{n}=(1-cd)^{\frac{n}{1-c}},
\end{equation}
where $0<c<1$, $0\leq d\leq \frac{1}{2}$. With the communication running,
Eve's successful detection probability decreases exponentially.

Summary, in this paper, we give an improved $ping-pong$ $protocol$.
Comparing with the $ping-pong$ protocol showed by Bostrom and Felbinger [5],
we use a single photon source instead of an EPR pair source. A stable and
efficient single photon source has been reported [10]. Knill, Laflamme, and
Milburn have shown that quantum logical operation can be performed using
linear optical elements and ancilla photons [11]. This secure communication
protocol can be carried out with great efficiency and speed using today's
technology.

This work was supported by the National Science Foundation of China (Grant
No. 10004013).

\section{References:}

[1]. C. H. Bennett and G. Brassard, 1984, in $proceedings$ $of$ $the$ $IEEE$ 
$International$ $Conference$ $on$ $Computers$, $Systems$ $and$ $%
\mathop{\rm Si}%
gnal$ $\Pr oces\sin g$, Bangalor, India, (IEEE, New York), pp. 175-179.

[2]. A. Ekert, Phys. Rev. Lett. 67, 661 (1991).

[3]. D. Bruss, Phys.Rev. Lett. 81, 3018 (1998).

[4]. A. Beige, B.-G. Englert, C. Kurtsiefer, and H. Weinfurter, Acta Phys.
Pol. A 101, 357 (2002).

[5]. Kim Bostr$\stackrel{..}{o}$m and Timo Felbinger, Phys. Rev. Lett. 89,
187902 (2002).

[6]. E. H. Lieb and M. B. Ruskai. Phys. Rev. Lett., 30, 434-436, (1973).

[7]. E. H. Lieb and M. B. Ruskai. J. Math. Phys., 14, 1938-1941, (1973).

[8]. M. A. Nielsen and I. L. Chuang. Quantum Computation and Quantum
Information (Cambridge University Press, Cambridge, UK, 2000).

[9]. W. F. Stinespring, Proc. Am. Math. Soc. 6, 211 (1955).

[10]. A. Beveratos, R. Brouri, T. Gacoin, A. Villing, J.-P. Poizat, and P.
Grangier. Phys. Rev. Lett. 89, 187901, (2002).

[11]. E. Knill, R. Laflamme, and G. J. Milburn. Nature (London) 409, 46
(2001).

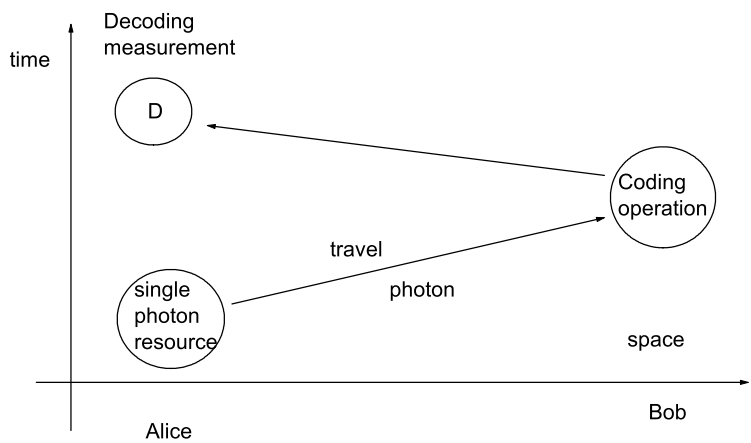
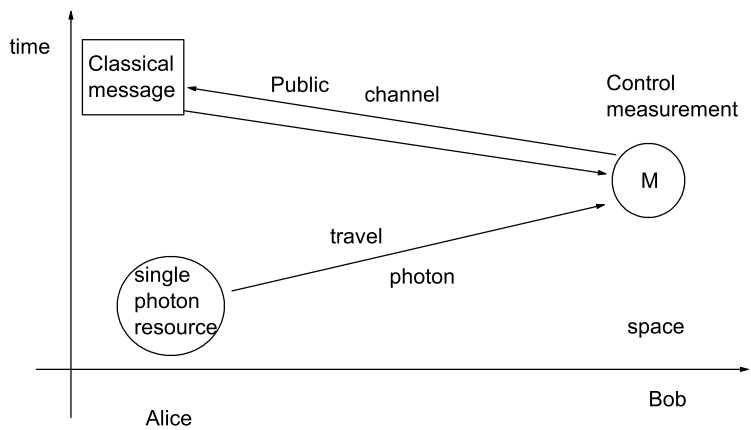
\end{document}